\documentclass[twocolumn,showpacs,preprintnumbers,amsmath,amssymb]{revtex4}

\usepackage{graphicx}
\usepackage{dcolumn}
\usepackage{bm}

\begin{document}


\title{Antiferromagnetic ordering and disappearance of pseudogap within the vortex core of Tl$_2$Ba$_2$CuO$_{6+\delta}$}

\author{K.~Kakuyanagi$^1$, K.~Kumagai$^1$, Y.~Matsuda$^2$, and M.~Hasegawa$^3$}
\affiliation{$^1$Division of Physics, Graduate School of Science, Hokkaido
University,  Sapporo 060-0810, Japan}%
\affiliation{$^{2}$Institute for Solid State Physics, University of Tokyo,
Kashiwanoha 5-1-5, Kashiwa, Chiba 277-8581, Japan}%
\affiliation{$^{3}$Institute of Material Research, Tohoku University, Katahira,
Sendai, 980-8577, Japan}%

\date{\today}

\begin{abstract}

	Spatially-resolved NMR is used to probe the magnetism in and around the vortex core of nearly optimally-doped Tl$_2$Ba$_2$CuO$_{6+\delta}$ ($T_c$=85~K).  The NMR relaxation rate $T_1^{-1}$ at $^{205}$Tl site, at which antiferromagnetic (AF) fluctuation can be monitored sensitively,  provides a direct evidence that the AF spin correlation is significantly enhanced in the vortex core region.   In the core region Cu spins show a local AF ordering  with moment $\sim 0.1\mu_B$ parallel to the layers at $T_N$=20~K.   Above $T_N$ the core region is in the paramagnetic state which is a reminiscence of the state above the pseudogap temperature ($T^*\simeq$120~K), indicating that the pseudogap disappears within the core.

\end{abstract}


\maketitle

 The relation between superconductivity and magnetism has been a central issue in the physics of high temperature superconductors (HTSC).  In HTSC the superconductivity appears when carriers are doped into the mother compounds, which are antiferromagnetic (AF) Mott insulators.   It is well established that the strong AF fluctuation plays a crucial role in determining many physical properties of the normal state of HTSC.  Recently, the influence of the AF fluctuation in the {\it superconducting} state has been attracting much attention.  In particular, how the antiferromagnetism emerges when the $d$-wave superconducting order parameter is suppressed is a fundamental problem in HTSC \cite{so5,ddw,sachdev}.  In this respect, the microscopic structure of the vortex core, which is a local normal region created by destroying the superconductivity by magnetic field, turns out to be a very interesting subject, especially since many unexpected behaviors have been observed experimentally.

  In conventional $s$-wave superconductors, the quasiparticles (QPs) are confined by the isotropic pair potential and form the bound states of Caroli, de Gennes and Matricon \cite{caroli}.  However the core structure of HTSC with $d_{x^2-y^2}$-wave symmetry is expected to be very different from that of ordinary superconductors, because the pair potential goes to zero at certain crystal directions. At an early stage, the vortex core structure of HTSC had been discussed within the framework of the ''semiclassical" approximation, which is a direct extension of the $s$-wave vortex core structure \cite{ichioka}. However, recent high resolution STM experiments have revealed many unexpected properties in the spectrum of the vortex core, which are fundamentally different from these semiclassical $d$-wave vortex cores \cite{stm}.  
A new class of theories have emphasized the importance of the magnetism arising from the strong electron correlation for accounting for the microscopic vortex core structure \cite{himeda,arovas,franz,han,lee,wang}.   Therefore it is crucial for the comprehension of  the vortex state of HTSC to clarify how the AF correlation and pseudogap, which characterize the magnetic excitation in the normal state, appear in and around the vortex core.

	Despite extensive studies, little is known about the microscopic electronic structure of the vortices, especially concerning the magnetism. The main reason for this is that although STM experiments can probe the local density of states (DOS) with atomic resolution, they do not directly reflect the magnetism.  Recent neutron scattering experiments on La$_{2-x}$Sr$_x$CuO$_4$ have reported that an applied magnetic field enhances the AF correlation in the superconducting state \cite{lake,lake2}.  These results were interpreted in terms of the competition between superconductivity and static AF ordering (or spin density wave, SDW)  \cite{sachdev}.  However, the relation between the observed AF static ordering and the magnetic excitations within the vortex core is still not clear, because the neutron experiments lack spatial resolution. Recent experimental \cite{curro,mitrovic,kakuyanagi} and theoretical \cite{takigawa,wortis,morr} NMR studies have established that the frequency dependence of spin-lattice relaxation rate $T_1^{-1}$ in the vortex state serves as a probe for the low energy excitation spectrum which can resolve {\it different spatial regions of the vortex lattice}. Up to now, however, all of these spatially-resolved NMR measurements have been carried out at the $^{17}$O sites \cite{curro,mitrovic,kakuyanagi}, at which the AF fluctuations are filtered, because the O atoms are located in the middle of neighboring Cu atoms with antiparallel spins \cite{masashi}.

	In this Letter we provide local information on the AF correlation in the different regions of the vortex lattice extending the measurements to the vortex core region, by performing a spatially resolved NMR imaging experiments on $^{205}$Tl-nuclei in nearly optimally-doped Tl$_2$Ba$_2$CuO$_{6+\delta}$. This attempt is particularly suitable for the above purpose because $T_1^{-1}$ at the Tl-site, $^{205}T_1^{-1}$, can monitor AF fluctuations sensitively. Quite
generally, $1/T_1$ is expressed in terms of the dynamical susceptibility as $\frac{1}{T_1}=\frac{\gamma_nk_BT}{2\mu_B^2}\sum_{q}|A_q|^2\frac{Im\chi(q,\omega_
0)}{\omega_0}$, where $\gamma_n$ is the nuclear gyromagnetic ratio, $A_q$ is the hyperfine
coupling between nuclear and electronic spins, and $\omega_0$ is the Larmor
frequency.  Because Tl atoms are located just above the Cu atoms and there exist large transferred hyperfine interactions between Tl and Cu nuclei through apical oxygen (see the inset of Fig.~1), Tl sees the full wavelength spectrum of Cu magnetic spin fluctuation; $^{205}T_1$ is dominated by $\chi$({\boldmath $q$}) at {\boldmath $q$} $=(\pi,\pi)$, {\it i.e.} AF fluctuations. This should be contrasted to the O-sites at which $\chi$({\boldmath $q$}) is dominated by uniform fluctuations at {\boldmath $q$}$=(0,0)$.  On the basis of the NMR imaging, we have been able to establish a clear evidence of the AF vortex core state of HTSC.

\begin{figure}
\includegraphics [scale=0.33,angle=-90] {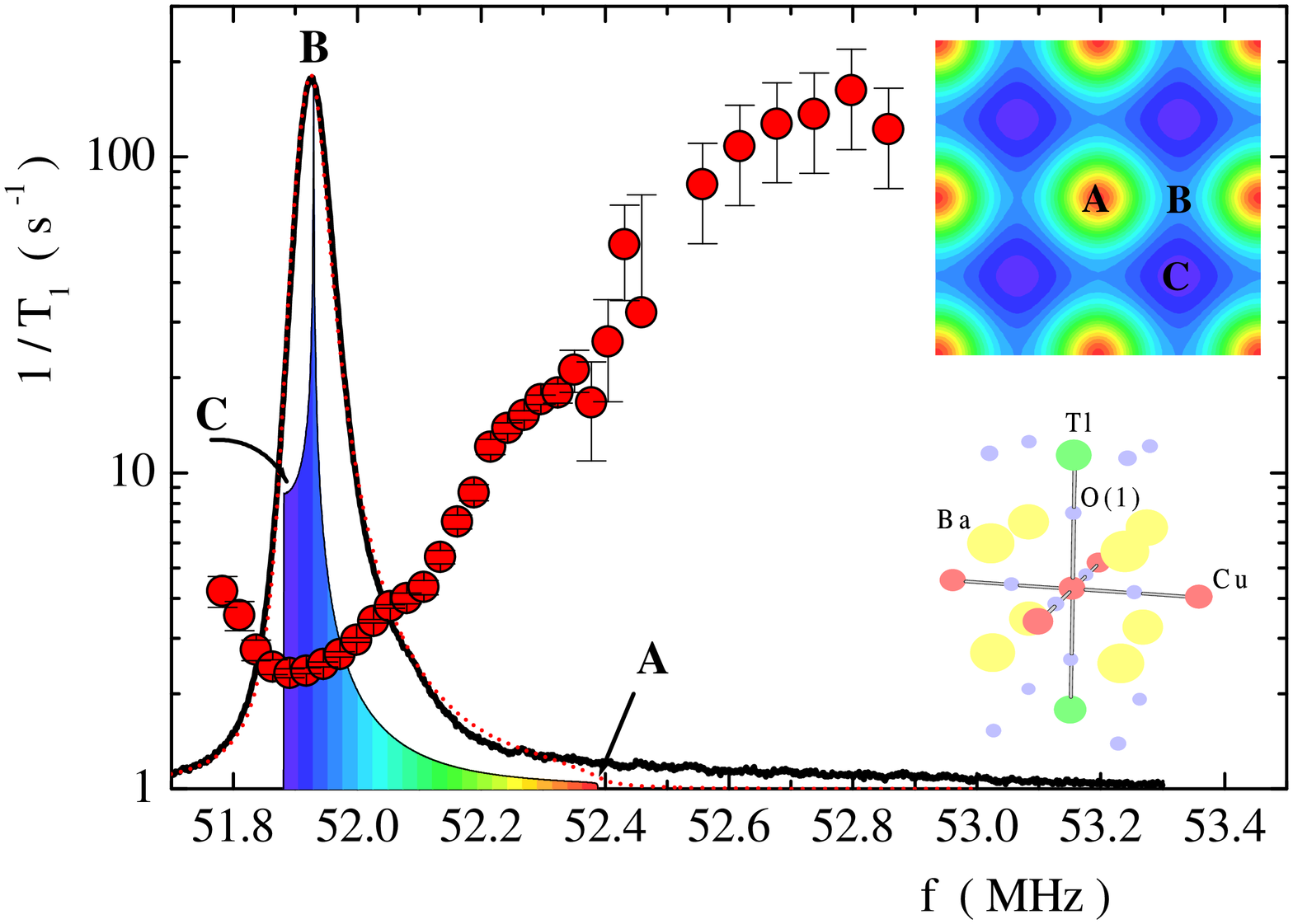}
\caption{$^{205}$Tl-NMR spectrum (solid line) at 5~K.  The thin solid line depicts the histogram at particular local fields obtained from Eq.(1).  The red dotted line represents the simulation spectrum convoluted with Lorentzian broadening function.  The red filled circles show the frequency dependence of $T_1^{-1}$  at the Tl site,  $^{205}T_1^{-1}$, at 5~K.  For details, see the text.  The upper inset shows the image of the field distribution in the vortex square lattice; vortex core (A), saddle point (B) and center of vortex lattice (C).   The lower inset shows the crystal structure of Tl$_2$Ba$_2$CuO$_{6+\delta}$. }
\end{figure}

	 NMR measurements were carried out on the $c$-axis oriented polycrystalline powder of high quality Tl$_2$Ba$_2$CuO$_{6+\delta}$ ($T_c$=85~K) in the external field along the $c$-axis. The $^{205}$Tl spin echo signals were obtained by a pulse NMR spectrometer.  A very sharp spectrum ($\sim$ 50~kHz) above $T_c$ was observed.  The spectrum becomes broad below $T_c$ due to the development of vortices.   The solid line in Fig.~1 depicts the NMR spectra at 5~K measured under the field cooling condition (FCC) in a constant field ($H_0$=2.1~T).  We stress here the importance of measuring  under FCC because the Bean critical current associated with the sweeping $H$ not only produces a field gradient in the crystal but also seriously influences $T_1$ by producing a shift to the QP energy spectrum.  The spectra was obtained by convolution of the respective Fourier-transform-spectra of the spin echo signals measured with an increment of 50~kHz.  A clear asymmetric pattern of the NMR spectrum, which originates from the local field distribution associated with the vortex lattice (the Redfield  pattern), is observed below the vortex lattice melting temperature ($\sim$60~K at $H_0$).   The local field profile in the vortex state is given by approximating $H_{loc}(\mbox{\boldmath$r$})$ with
the London result,
\begin{equation}
H_{loc}(\mbox{\boldmath$r$})=H_0\sum_{G}\exp{(-i\mbox{\boldmath$G\cdot
r$})}~\frac{\exp({-\xi_{ab}^2\mbox{\boldmath$G$}^2/2})}
{1+\mbox{\boldmath$G$}^2\lambda_{ab}^2},
\end{equation}
where $\mbox{\boldmath$G$}$ is a reciprocal vector of the vortex lattice, $\mid\mbox{\boldmath$r$}\mid$ the distance from the center of the core, $\xi_{ab}$  the in-plane coherence length, and $\lambda_{ab}$  the in-plane penetration length.   The thin solid line  in Fig.~1 depicts the histogram at a particular local field which is given by the local field distribution $f(H_{loc})=\int_{\Omega} \delta [H_{loc} (\mbox{\boldmath$r$})-H_{loc}]d^2\mbox{\boldmath$r$}$ in Eq.(1), where $\Omega$ is the magnetic unit cell. In the calculation we used $\xi_{ab}$=18\AA~and $\lambda_{ab}=1700~\AA$, and assumed the square vortex lattice. The upper inset shows the image of the field distribution in the vortex lattice. The magnetic field is lowest at the center of the vortex square lattice (C-point), and it is highest at the center of the vortex core (A-point).  The intensity of the histogram shows a peak at the field corresponding to the saddle point  (B-point).  The real spectrum broadens due to the imperfect orientation of the power. The red dotted line represents the spectrum convoluted with Lorentzian broadening function,
$f(H_{loc})=\sigma/(4H_{loc}^2 + \sigma^2)$ using $\sigma$=48kHz.  The theoretical curve reproduces the data well except at the high frequency region where deviations become significant.  We will discuss this high frequency tail later.

\begin{figure}
\includegraphics [scale=0.30,angle=-90] {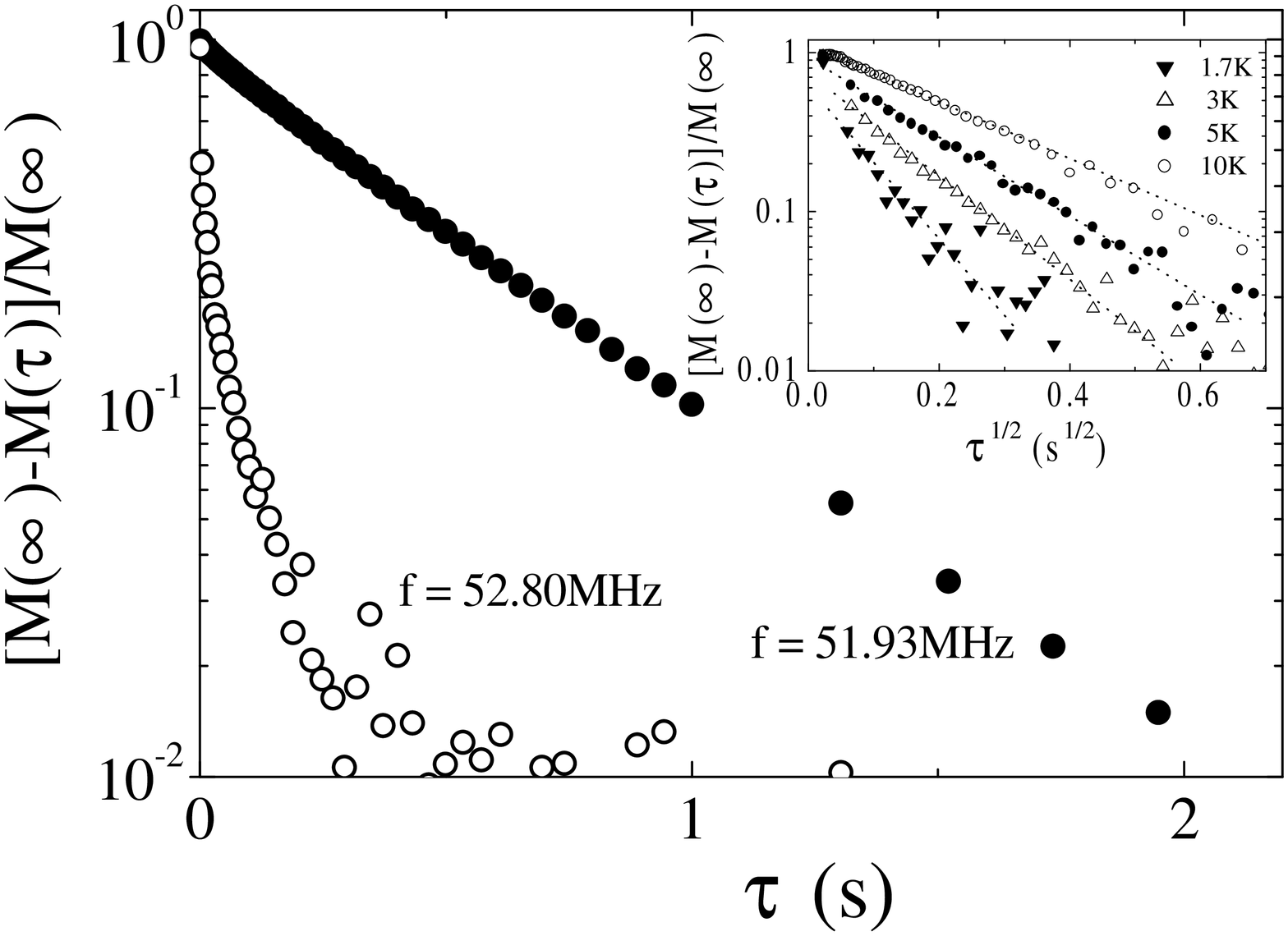}
\caption{Recovery curves of nuclear magnetization of $^{205}$Tl  as a function of $\tau$.   The filled circles represent the data at the saddle point (B-point in Fig.~1, $f$=51.93~MHz) and open circles represent the data at the vortex core (A-point in Fig.~1, $f$=52.80MHz), respectively.  The inset shows the same data as a function of $\sqrt \tau$ at several temperatures at the vortex core.  }
\end{figure}

	The spectrum shown in Fig.~1 demonstrates that the NMR frequency depends on the position of the vortex lattice. Therefore we can obtain the spatially-resolved information of the low energy excitation by analyzing the frequency distribution of the corresponding NMR spectrum.  For $^{205}$Tl with nuclear spin $I$=1/2, the recovery curve of the nuclear magnetization $M(t)$ fits well to a single exponential relation, $R(\tau)=(M(\infty)-M(\tau))/M(\infty)= \exp(-\tau/T_1)$ in the normal state. In the vortex state, on the other hand, the feature of the recovery curves is strongly position dependent. Figure 2 and its inset display the recovery curves as a function of time, $\tau$, at the saddle point (B-point in Fig.~1, filled circles) and at the vortex core (A-point in Fig.~1, open circles).  The procedure to determine $T_1$ is as follows. The spin echo intensities are measured as a function of $\tau$ after saturation pulses. Then the nuclear magnetization recovery curves as a function of $\tau$ is obtained from each frequency component of the Fourier transform spectra.  We obtained the data set of the recovery intensity for each frequency point at the 28~kHz interval with a gaussian weight function of $\sigma$=10~kHz. There are two distinct features. First, the decay time at the center of the core is much faster than at the saddle point. Second, while the recovery curves show the single exponential at the saddle point, they show a $\sqrt{\tau}$ dependence at the core region.  We will discuss this $\sqrt{\tau}$ dependence later.   In what follows, we defined $T_1$ as the time required for the nuclear magnetization to decay by a factor 1/$e$, in order to define $T_1$ uniquely for either decay curve.
		
\begin{figure}[b]
\includegraphics [scale=0.30,angle=-90] {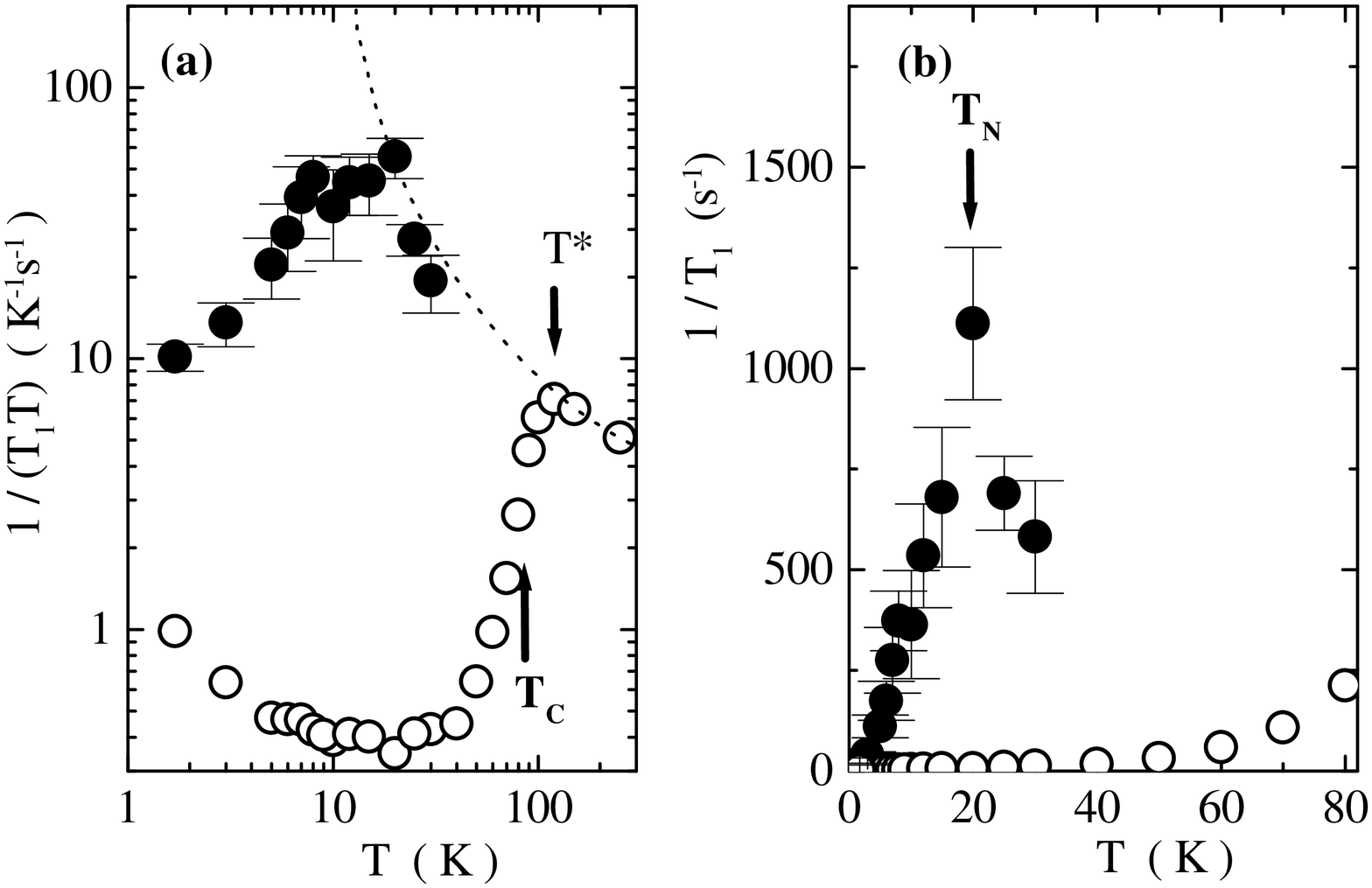}
\caption{ $T$-dependence of $(^{205}T_1T)^{-1}$ (a) and  $^{205}T_1^{-1}$ (b).  The filled  circles represent the data at the vortex core (A-point in Fig.~1) and open circles represent the data at the frequency correspond to the saddle point (B-point in Fig.~1).  In (a) $T^*\simeq 120$~K is the pseudogap temperature. The dotted line represents the Curie-Weiss law which is determined above $T^*$. In (b), $T_N$ is the temperature at which $^{205}T_1^{-1}$ at the core, $(^{205}T_1^{core})^{-1}$, exhibits a sharp peak, which turns out to be the N\'eel temperature within the vortex core.    }
\end{figure}
		
	The red filled circles in Fig.~1 show the frequency dependence of $^{205}T_1^{-1}$. On scanning from the outside into the core (C$\rightarrow$B$\rightarrow$A), $^{205}T_1^{-1}$ increases rapidly after showing a minimum near the saddle point (B).  The magnitude of $^{205}T_1^{-1}$ in the core region (A) is almost two orders of magnitude larger than that near the saddle point (B). This large enhancement of $^{205}T_1^{-1}$ is in striking contrast to $^{17}T_1^{-1}$ at $^{17}$O sites reported in YBa$_2$Cu$_3$O$_7$ \cite{curro,mitrovic} and  YBa$_2$Cu$_4$O$_8$ \cite{kakuyanagi}, in which the enhancement of $^{17}T_1^{-1}$ at the core region is 2-3 times at most  and has been attributed to local DOS produced by a Doppler shift of the QP energy spectrum by supercurrents around the vortices \cite{volovik}.  Therefore, the remarkable enhancement of $^{205}T_1^{-1}$ provides a direct evidence that {\it the AF correlation is strongly enhanced near the vortex core region.} The decrease of $^{205}T_1^{-1}$ well outside the core when going from point C to B was also reported in $^{17}T_1^{-1}$ in YBa$_2$Cu$_3$O$_7$\cite{mitrovic} and YBa$_2$Cu$_4$O$_8$ \cite{kakuyanagi}.  This phenomenon has been discussed in terms of the suppression of the AF fluctuations by the Doppler shifted QP DOS \cite{kakuyanagi,morr}.	
\begin{figure}[b]
\includegraphics [scale=0.30,angle=-90] {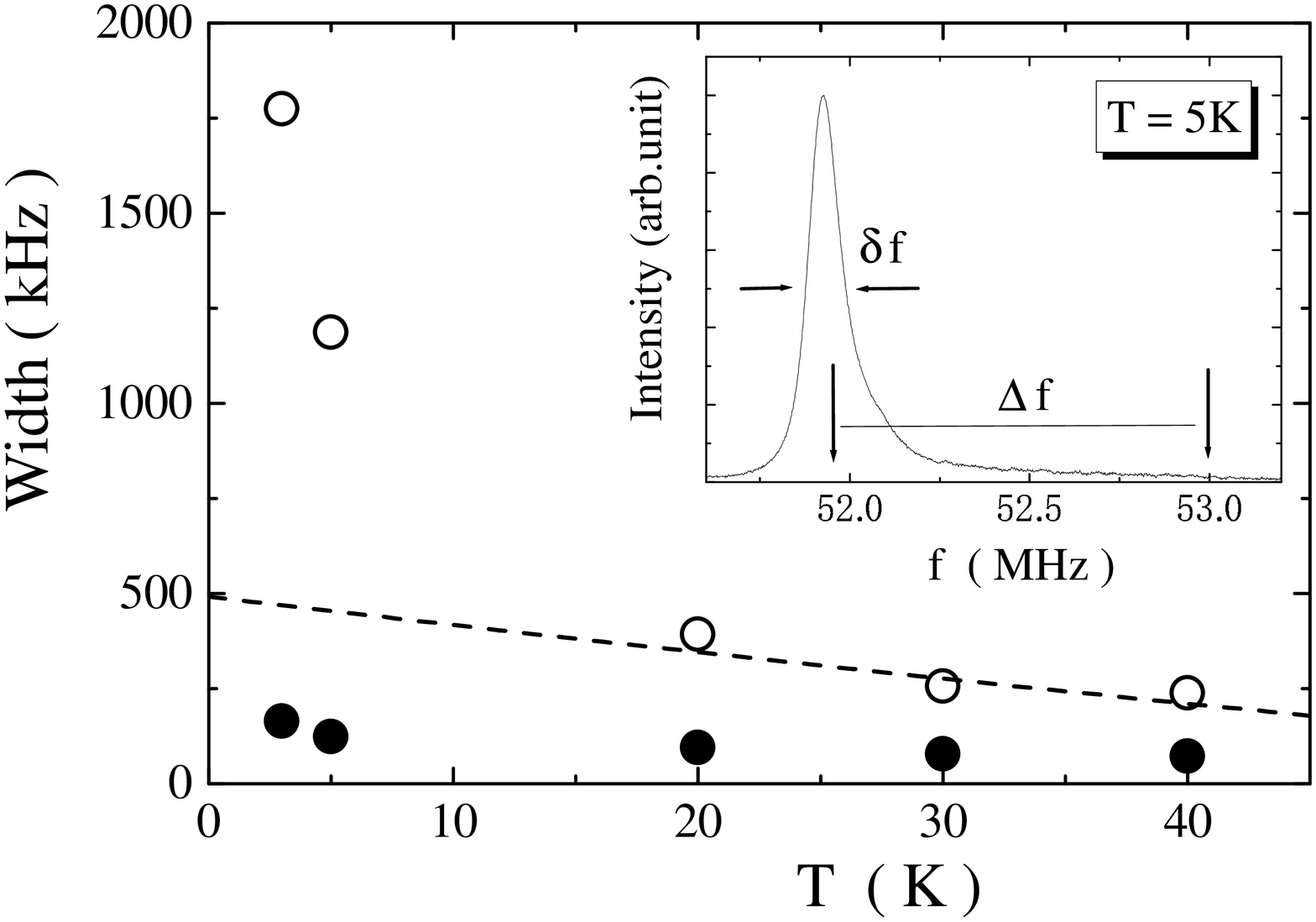}
\caption{ The line widths of the $^{205}$Tl-NMR spectrum plotted as a function of temperature. $\Delta f$ (open circles) indicates the line width determined by the frequency at which the intensity becomes 1\% of the peak. The dashed line represents the line width calculated from the Redfield pattern (frequency difference between A- and B- point in Fig.~1).  $\Delta f$ increases rapidly below $T_N=20$~K.  Filled circles represent $\delta f$ which is defined as the line width at the half intensity.  Inset shows the spectrum at 5~K and the  definition of $\Delta f$ and $\delta f$. }
\end{figure}

	Figures 3 (a) and (b) depict the $T$-dependences of $(^{205}T_1T)^{-1}$ and $^{205}T_1^{-1}$ within the core (filled circles) and at the frequency corresponding to the saddle point (open circles).  We comment on here the influence of vortex vibration on $^{205}T_1$ discussed in Ref.\cite{VV}.  Generally  $^{205}T_1^{-1}$ is composed of contributions from vortex vibration, $^{205}T_{1,VV}^{-1}$, and  from the magnetic excitation $^{205}T_{1,ME}^{-1}$; $^{205}T_{1}^{-1}=^{205}T_{1,VV}^{-1}+^{205}T_{1,ME}^{-1}$.   However, $^{205}T_1^{-1}$ in the core region, $(^{205}T_1^{core})^{-1}$,  is two orders of magnitude larger than that expected solely from vortex vibration at all temperatures (see Fig.~2 in Ref.\cite{VV}).  Therefore  the influence of the vortex vibration is negligibly small in the core region.    From high temperatures down to about 120~K,  $(^{205}T_1T)^{-1}$ obeys the Curie-Weiss law, $(^{205}T_1T)^{-1}\propto1/(T+\theta)$. The lowest $T$ at which this law holds is conveniently called the pseudogap temperature $T^*$.   Below $T^*$,  $(^{205}T_1T)^{-1}$ decreases rapidly without showing any anomaly associated with the superconducting transition at $T_c$, similar to other HTSC \cite{masashi}. Below 40K, $(^{205}T_1T)^{-1}$ is nearly $T$-independent down to 4~K due to the DOS induced by the impurity in $d$-wave superconductor. The $T$-dependence of $(^{205}T_1^{core})^{-1}$ contains some key features for understanding the core magnetism.  

	The first  important signature is that $1/^{205}T_1^{core}$ exhibits a sharp peak at  $T$=20~K, which we label as $T_N$ for future reference (Fig~3(b)).    Below $T_N$ $(^{205}T_1^{core})^{-1}$ decreases rapidly with decreasing $T$.   There are two possible origins for this peak.  One is the reappearance of the pseudogap and the other is the occurrence of a local static AF ordering (or local SDW ordering) in the core region.  The sharp peak of $^{205}T_1^{-1}$ at $T_N$ seems to support  AF ordering.   The broadening of the Redfield pattern at the high frequency region discussed before also gives an additional evidence on  AF ordering. The open circles in Fig.~4 display the line width at the high frequency tail, $\Delta f$, which is defined as a difference between the frequency at the peak intensity  and the frequency at which the intensity becomes 1\% of the peak.  For the comparison, we plot  the line width calculated from the Redfield pattern  (frequency difference between A- and B- point in Fig.~1) by the dashed line.  At high temperatures $\Delta f$ agrees well with the calculation, while below $T_N$=20~K it becomes much larger.  We also plot $\delta f$ which represents the line width at the half intensity (filled circles).  Since $\delta f$  changes little below $T_N$, the line broadening occurs only near the core region.  We stress that the high frequency tail below $T_N$ is naturally explained by  the transferred hyperfine fields at the Tl site through apical oxygen induced by the AF ordering within the core, which causes additional broadening. We also point out that the appearance of the local AF ordering is also consistent with the $\sqrt{\tau}$ dependent nuclear magnetization decay curve shown in the inset of Fig.~2.  In fact, the $\sqrt{\tau}$ dependence has been observed when the microscopic imhomogeneous distribution of $T_1^{-1}$ due to strong magnetic scattering centers is present \cite{sil}.  On basis of these results, we are lead to conclude that the vortex core region  shows local AF ordering at $T_N$=20~K; $T_N$ {\it corresponds to the N\'eel temperature within the core.} This AF ordering within the core is  consistent with the prediction of recent theories based on the $t-J$ and SO(5)  models \cite{himeda,arovas,franz,han}. 

	The second important signature for the core magnetism is that, as shown by the dotted line in Fig.~3(a),  $(^{205}T_1^{core}T)^{-1}$ above $T_N$ nearly lies on the Curie-Weiss law line extrapolated above $T^*$.  This fact indicates that the vortex core region appears to be {\it in the paramagnetic state which is a reminiscence of the state above $T^*$; the pseudogap is absent in the core region.}    We note that the present results seem to be inconsistent with the recent theories which predict  local orbital currents, in which the pseudogap phenomenon within the core is assumed \cite{lee,wang}.  The present results also should be distinguished from those of the neutron scattering experiments on La$_{2-x}$Sr$_x$CuO$_4$ \cite{lake}, in which the static SDW coexists with superconductivity even in zero field just below $T_c$.  In the present compound, on the other hand, we do not observe such a static SDW ordering and the vortex core region is in the paramagnetic state in a wide $T$-region between $T_c$ and $T_N$.

	We finally discuss the spin structure within the core. The broadening occurs only at high frequencies while it is absent at low frequencies. This fact indicates that {\it the AF spins are oriented parallel to the CuO$_2$ layers}. This follows by observing that the broadening should occur at both high and low frequency sides if the AF ordering occurs perpendicular to
the layers, because in this case the direction of the alternating transferred hyperfine fields are parallel and antiparallel to the applied field. Using the hyperfine coupling constant, $A_{hf}$= 56kOe/$\mu_B$, the magnetic moments induced within the core is estimated to be $\sim$0.1$\mu_B$.

       Summarizing the salient features of spatially-resolved NMR results in the vortex lattice; (1) Upon approaching the vortex core, $(^{205}T_1)^{-1}$ is strongly enhanced (Fig.~1),  (2) Near the core region, the NMR recovery curves show the $\sqrt{\tau}$-dependence (Fig.~2),  (3) $(^{205}T_1^{core})^{-1}$ shows a peak at $T$=20~K (Fig.~3),  (4) NMR spectrum near the core region broadens below $T$=20~K (Fig.~4).  All of these results provide  direct evidence that in the vortex core region the AF spin correlation is extremely enhanced, and that the paramagnetic-AF ordering transition of the Cu spins takes place
at $T_N=20$~K.  
We also find the pseudogap disappears within the core.  The present results offer a new perspective on how the AF vortex core competes with the $d$-wave superconductivity.

	We acknowledge helpful discussions with M.~Franz, M.~Imada, J.~Kishine, K.~Machida, D.K.~Morr, M.~Ogata, S.H.~Pan, M.~Takigawa, A.~Tanaka, and Z.~Tesanovi\'c.


\end{document}